\magnification=\magstep1
\tolerance 500
\rightline{TAUP 2309-95}
\rightline{27 December, 1995}
\vskip 5 true cm
\centerline{\bf HYPERCOMPLEX QUANTUM MECHANICS}
\bigskip
\centerline{L.P. Horwitz}
\centerline{School of Physics}
\centerline{ Raymond and Beverley
Sackler Faculty of Exact Sciences}
\centerline{Tel Aviv University, Ramat Aviv 69978, Israel}
\centerline {and}
\centerline{Department of Physics}
\centerline{Bar Ilan University, Ramat Gan 52400, Israel}
\vskip 3 true cm
\noindent {\it Abstract\/} The fundamental axioms of the quantum
theory do not explicitly identify the algebraic structure of
the linear space for which orthogonal subspaces correspond to
the propositions (equivalence classes of physical questions).
The projective geometry of the weakly modular orthocomplemented
lattice of propositions may be imbedded in a complex Hilbert
space; this is the structure which has traditionally been used.
This paper reviews some work which has been devoted to
generalizing the target space of this imbedding to Hilbert
modules of a more general type. In particular, detailed discussion
is given of the simplest generalization  of the complex Hilbert
space, that of the quaternion Hilbert module.
\smallskip
\noindent{\it Key Words:\/} Octonions, Clifford Algebras, quantum
theory, quaternion Hilbert modules, tensor product.
\vfill
\eject
\noindent {\bf 1. Introduction}
\par In his discussion of the development of the theory of matrices
in the middle of the nineteenth century, in which he remarked
that ``it seems almost uncanny how mathematics now prepared
itself for its future service in quantum mechanics,'' Max
Jammer$^{1}$
recounted how the natural generalization of the real numbers to
complex numbers and quaternions played a central role.  He cites
Tait$^{2}$  as attributing to Hamilton the discovery of matrices
in a letter to A. Cayley, who discovered that quaternions could
be represented as $2 \times 2$ matrices over complex elements.
Hamilton$^{3}$ invented the quaternions in 1844; Tait referred to
Hamilton's ``linear and vector operators,'' and called
Cayley's discovery only a modification of Hamilton's ideas.
Taber$^{4}$, in 1890, renewed the claim that Hamilton had indeed
originated the theory of matrices.  Gibbs$^{5}$ ``regarded
1844 as a `memorable' year in the annals of mathematics because
it was the year of the appearance of Hamilton's first paper$^{3}$ on
quaternions.''$^{1}$.
\par John von Neumann, in fact, emphasized
the result of Hurwitz$^{6}$, that there are just {\it four}
normed division algebras, the real ({\bf R}), complex ({\bf C}),
quaternion ({\bf H}), and octonion, or Cayley, algebra ({\bf O}),
and that ``nature must make use of them.''$^{7}$.
\par Herman Goldstine and I set out to investigate the possibility
of constructing a Hilbert space, with application to a more general
form of the quantum theory, using the Cayley numbers {\bf O}
as coefficients on the vector space (they are not commutative or
associative)$^8$.
\par In fact, the octonions arise in a natural way as a result of
the attempt of Jordan, von Neumann and Wigner$^{9}$ to set up
a quantum theory in which the products of observables (represented,
in general, by non-commuting self-adjoint operators on the complex
Hilbert space) remain observables.  They defined multiplication
as the symmetric product $A \circ B = {1 \over 2}(AB + BA)$. This
definition gives rise to a set of algebraic relations, and one may
ask for the solution to the inverse problem, i.e., to ask for
which algebraic structures could these relations be valid.  The
answer, provided by Albert$^{10}$, was that only the symmetric
product of matrices on the real, or complex, are allowed, with
one exception, the $3 \times 3$ matrices over octonions.
  \par The lack of associativity formed an obstacle to us
in constructing the adjoint of operators on the Hilbert space.
We found, however, in the construction of the representation
of a vector in terms of a complete orthogonal set, that the
alternative property of the octonions admits the closure of the
subspace generated by (successively associated) products of the
vector with octonion elements to order {\it seven}, i.e., after
multiplication seven times by octonions, the subspace no longer
grows.  The algebra of successive multiplications is isomorphic
to the Clifford algebra of order seven.  We found, furthermore,
that the minimal ideals of this Clifford algebra reduce products
of Clifford elements into single elements, resulting in effective
multiplication laws that reproduce those of the non-associative
Cayley algebra.  We therefore turned to the general problem
of constructing Hilbert modules over Clifford algebras, in
particular, that of $C_7$, the Clifford algebra of order seven.
\par The automorphism group of the Clifford algebra $C_7$ which
stabilizes a minimal ideal is isomorphic to the
automorphism group of the octonions, i.e., $G_2$.  In 1965,
this result enabled Biedenharn and me to generalize the quark-lepton
model of G\"unaydin and G\"ursey$^{12}$, set in the framework
of an octonionic Hilbert space, to the Hilbert module over $C_7$
$^{13}$.
\par In the following, I discuss some properties of the octonionic
Hilbert space, and how this structure leads to a Clifford Hilbert
module.  In the succeeding sections, I specialize to the simplest
of the Clifford algebras beyond the complex, which is also a
division algebra, that of the quaternions.  In this work, spanning
many years of attention, the historical account and scientific
evaluation given in Max Jammer's book was a strong element of
encouragement.  It is a pleasure to dedicate this review to him
on the occasion of his eightieth birthday.
\bigskip
\noindent {\bf 2. Octonionic and Clifford Modules}
\smallskip
\par To see how the Clifford module structure emerges, consider the
definition$^8$ of the octonion Hilbert space.
If $f,g \in {\cal H}$, we suppose that $fa + gb \in {\cal H}$,
$(f,f) = \Vert f \Vert^2 \geq 0$ ($f=0$
 in case of the equality), and the
linearity property $(fa,fa) = \Vert f\Vert^2 |a|^2$, where the
constants $a,b$ , of the form $\sum_{i=0}^{7} \lambda_i
e_i$, for $\{\lambda_i\}$ real, are elements of the involutive
$(e_i ^* = -e_i , i \neq 0),  e_ie_j =- e_je_i, e_i^2 = -1
({\rm non-zero}\ \  i \neq j), e_1e_2 = e_3, e_5 e_1 = e_4, e_4 e_2
= e_6, e_6 e_3 = e_5, e_6 e_7 = e_1, e_5 e_7 = e_2, e_4 e_7 = e_3
({\rm and\  cyclic}), e_0 = 1$.  These are seven (associative)
quaternion subalgebras.  The non-associativity follows from the
relation $e_i(e_j e_k) = - (e_i e_j)e_k $ for $i,j,k$ not all in
one of the quaternion subalgebras.  Assuming that the space is
separable, one may construct an orthonormal set of vectors
$\{\varphi_n\}$ which generate subspaces $\{\varphi_n a_n \}$
that span the space$^8$.  Orthogonalization is carried out by
a variational principle, by means of which we can construct,
 for any vector
$g$, a part in $\{fa \vert a \in {\bf O}\}$, and a part
orthogonal to this manifold. The orthogonal part is found by
imposing the requirement that $g-fa = h$ be minimum (in norm),
and hence $\Vert h + fb \Vert \geq \Vert h \Vert$.  With this,
one obtains $ {\rm Re} (h,fb) = 0$ for any $b$.  The factor
$b$ may be extracted under the real part$^{14}$, so that
$ {\rm Re}[(h,f)b] = 0, {\rm or} (h,f) =0$.  But, if we
expand some arbitrary vector $g$ in terms of the orthogonal
expansion $\sum \varphi_na_n$, in general
$(\varphi_m \varphi_n a_n) \neq 0, n \neq m$, so we cannot solve
for the coefficients $a_n$.  One may then attempt to widen the set
of subspaces to the form $(fa)b$; the orthogonalization can
be carried out in the same way.  But again, one cannot solve for the
coefficients in $ \sum (\varphi_n a_n)b_n$.
\par The algebra has, however, the alternative property$^{14}$,
for which $(ab)b = ab^2$, so one may assume $(fa)a = fa^2$ for
any $f \in {\cal H}$.  It follows that $(fa)b + (fb)a = f(ab+ba)$,
so that for the octonion elements, for example,
$(fe_1)e_2 + (fe_2)e_1 = 0$.  Hence the growth of the size of the
subspace generated by a vector $f$ is limited by the multiplication
, successively, by only seven octonion elements.  The operation of
successive multiplication $(\cdots((fa)b)\cdots c)$ is
necessarily associative, and the seven elements of the Cayley
algebra, with unity, generate in this way the algebra of the
seventh Clifford algebra $C_7$.
\par The elements of the Clifford algebra $C_7$ close on a group
of 256 elements; since (we now drop the associating parentheses)
$e_1 e_2 \cdots e_7$ commutes with all elements, and its square is
one, we can define the algebraic projections $P_{\pm}
= {1 \over 2}(1 \pm e_1 e_2 \cdots e_7)$.  Defining a new scalar
product in the module over $C_7$ so that $(f,g)
= (g,f)^\dagger \in C_7, {\rm Tr}(f,ga) = {\rm Tr} (fa^\dagger,g)$
( $C_7$ is isomorphic to a matrix algebra)$^{11}$, we see that the
(right-acting) algebraic projections $P_\pm$ are Hermitian
(this involution is carried to $e_i^\dagger = -e_i, i \neq 0$).
The norm is defined by $\Vert f\Vert^2 = {\rm Tr}(f,f)$.
We remark that transition probabilities in such a theory are
calculated by $P_{\psi \chi} = {\rm Tr} [(\psi,\chi)(\chi,\psi)]$,
if $\psi, \chi$ are normalized.
\par The Clifford projections
$$ P^\pm_0 = {1 \over 8} (1-e_1 e_2 e_3)
 (1- e_4 e_2 e_6) (1 - e_4 e_5 e_1) P_\pm  $$
are minimal ideals in each sector
 of the split representation.  The
elements $$e_1 e_2 e_3, e_4 e_2 e_6,\  {\rm and}\  e_4 e_5 e_1$$
commute with each other and are Hermitian.  The representation of
$C_7$ is reduced by $P_\pm$, and the $8 \times 8$ matrices in each
of the two invariant subspaces are carried algebraically by the
$e_{ij}^\pm = e_i P_0^\pm e_j^\dagger, i,j = 0, \dots 7$,
corresponding to matrices of all zero elements but for the $i,j$
component, which is unity.  It is remarkable that, as mentioned above,
for example,
$$ e_1 e_2 P_0^\pm = e_3 P_0^\pm.$$
The remaining {\it non-associative} multiplication laws of the
Cayley algebra reappear in this way as well (with a sign change
in those involving $e_7$ in the $\pm$ sectors).
\par Hence the automorphisms of $C_7$ that preserve $P_0^\pm$
are the same as those of the Cayley algebra, constituting the
group $G_2$.
\par We leave the generalized
structure of the $C_7$ module for discussion elsewhere,
 and concentrate in the following on the properties
of the simplest Clifford module beyond the complex, that of $C_2$,
equivalent to the quaternions (generated by $1 \equiv e_0, e_1,
e_2, {\rm and}\  e_3 \equiv e_1 e_2 $).
\bigskip
{\bf 3. Quaternionic Quantum Mechanics}
\smallskip
\par The quaternionic Hilbert space$^{15}$ is defined by a set of
elements $f \in {\cal H}$; if $f\in {\cal H}, g \in {\cal H},
fa + gb \in {\cal H}$, where $a,b \in {\bf H}$, the algebra
of quaternions defined as real linear combinations of the
$\{e_i\}$, where $e_i^2 = -1, e_1 e_2 e_3= -1, e_i^* = -e_i,
i \neq 0$, and $(f,g) = (g,f)^* \in {\bf H}$, with $(f,f) =
\Vert f \Vert^2 \geq 0 $ defining the norm.  Transition probabilities
are given by $P_{\psi \chi} = \vert (\psi,\chi)\vert^2$, which
coincides with the general form given above for Clifford algebras
since the quaternions are a normed division algebra as well.
\par The scalar product has the linearity property
$(f,ga) = (f,g)a , \forall a \in {\cal H}$. Such a Hilbert module can
be used to represent the quantum theory$^{15,16,17}$; this
structure contains the usual complex theory and its results,
but predicts new effects as well.
\par  In order to construct some physical observables in terms
of self-adjoint and anti-self-adjoint operators, let us define
the action of translation by means of the representation
of a state vector on the spectral resolution of the self adjoint
operator of position $X$ (the spectral theorem of von Neumann
is true in the quaternion Hilbert space$^{15,16}$) as
$$ \langle x+ \delta x \vert f) = \langle x \vert T(\delta x) f),
\eqno(3.1) $$
where $T(\delta x)$ is a unitary operator.  Expanding to lowest order
in $\delta x$, so that
$$ T(\delta x) = 1 + \delta x S + O(\delta x^2), \eqno(3.2)$$
it follows from $(3.1)$ that
$$ \langle x \vert Sf) = {\partial \over \partial x} \langle x
\vert f), \eqno(3.3)$$
where $S$ is the anti-self-adjoint operator
$$ S= \int \vert x \rangle {\partial \over \partial x} \langle
x\vert dx. \eqno(3.4)$$
If we define the {\it quaternion linear} operator (satisfying
$P(fa) = (Pf)a) $\footnote{*}{There are, as well, complex linear
operators $A(fz) = (Af)z, z \in {\bf C}(1,e), e^2 = -1$, but
for which $A(fa) \neq (Af)a$ in general, and operators that are
real linear only.}
$$ \eqalign{ P &= \hbar E_i S   \cr
&= -\hbar \int \vert x \rangle e_i
{\partial \over \partial x} \langle x
\vert dx , \cr} \eqno (3.5) $$
where $ E_i $ belongs to a {\it left} algebra isomorphic to
${\bf H}$,
$$ E_i = \int \vert x \rangle e_i \langle x \vert dx, \eqno (3.6)$$
then one finds that
$$ [X,P] = -\hbar E_i, \eqno(3.7)$$
and, by a proof somewhat more involved than that for the complex
Hilbert space$^{16}$,
$$ \Delta X \Delta P \geq \hbar / 2. \eqno(3.8)$$
\par The existence of such left algebras is important
(in the presence of
these left algebras, the vector space is often called a
 {\it bimodule}) for the construction of tensor
product spaces to represent many-body systems, as well as
in applications to the one particle problem; I recall here
a basic representation theorem$^{16, 18}$.  Let us define a
left acting algebra $\{E_i\}$ isomorphic to ${\bf H}$, for which the
operator norm is unity, i.e.,${\rm sup}\Vert E_i h \Vert =
\Vert E_i \Vert \Vert h \Vert = \Vert h \Vert$.  Then, for
any $g,f \in {\cal H}$, there is a unique decomposition
$$ f = \sum_i E_i f_i, \qquad g = \sum_i E_i g_i \eqno(3.9)$$
where $E_i f_i = f_i e_i, E_i g_i = g_i e_i$, for which
$(f_i,g_j)$ is real.  We call the vector valued coefficients
$\{f_i \}$ ``formally real''.  The operators $\{E_i \}$ may be of
the form $(3.6)$, but there are, in principle, an infinite number
of such algebras.
\par It is interesting to examine the structure of the Schr\"odinger
equation in this context.  As Adler$^{17}$ has shown, the time
independent Schr\"odinger equation implies the existence of
an ``optical potential'' which can break time reversal invariance in
a very natural way.  Consider the time dependent Schr\"odinger
equation
$$ {\partial \psi \over \partial t} = - {\tilde H} \psi,
\eqno(3.10) $$
where $\psi$ is a quaternion-valued $L^2$ function and
${\tilde H}$ is an anti-self-adjoint Hamiltonian operator.  The
stationary problem corresponds to
$$ {\tilde H} \psi = \psi e E , \eqno(3.11)$$
where $E$ is real, and $e$ is an imaginary quaternion (as required
by the anti-self-adjoint property of ${\tilde H}$).  Multiplying
$(3.11)$ by the quaternion $q$ on the right, we obtain
$$ {\tilde H} \psi q = \psi q (q^{-1}e q) E, \eqno(3.12)$$
and by an appropriate choice of $q$, the imaginary unit $e$
can be brought, say to $e_1$, and the sign of $E$ assumed
positive.  The standard form of $(3.11)$ can therefore be written
as
$$ {\tilde H} \psi = \psi e_1 E, \qquad E \geq 0. \eqno(3.13)$$
\par Now, every quaternion $$a= a_0 + \sum_{i=1}^{3} a_i e_i  \qquad
(\{a_i\}{\rm real})$$
can be written as
$$ a = a_\alpha + e_2 a_\beta , \eqno(3.14)$$
where $a_\alpha = a_0 + a_1 e_1, a_\beta = a_2 - a_3 e_1 $, so
that
$$ \psi(x) = \psi_\alpha(x) + e_2 \psi_\beta(x), \eqno(3.15)$$
where
$$\eqalign{ \psi_\alpha(x) &= \psi_0(x) + e_1 \psi_1(x) \cr
            \psi_\beta(x) &= \psi_2 (x) - e_1 \psi_3(x). \cr }
\eqno(3.16)$$
\par The relations $(3.15), (3.16)$ are the $x$-representation
of $$ \eqalign{ \psi &= \psi_\alpha + E_2 \psi_\beta \cr
          \psi_\alpha &= \psi_0 + E_1 \psi_1 \cr
          \psi_\beta  &= \psi_2 + E_1 \psi_3 \cr} \eqno(3.17)$$
in the abstract quaternionic Hilbert space, where the $\{E_i\}$
are of the form $(3.6)$.
\par If ${\tilde H}$ is quaternion linear, it also has the
decompositon (in $x$-representation, as in $(3.15)$)
$$ {\tilde H} = H_\alpha + e_2 H_\beta = -{\tilde H} ^\dagger
\eqno(3.18)$$
where $ H_\alpha = e_1 H_1, H_1^\dagger = H_1$.  Decomposing
$(3.13)$ by components, one finds$^{17}$
$$ \eqalign{
e_1 H_1 \psi_\alpha - H_\beta^* \psi_\beta &= e_1 E \psi_\alpha
\cr -e_1 H_1^* \psi_\beta + H_\beta \psi_\alpha
&= e_1 E \psi_\beta, \cr} \eqno(3.19)$$
where the asterisk indicates complex conjugation ($e_1 \rightarrow
-e_1$) in the complex subalgebra ${\bf C}(1,e_1)$ of ${\bf H}$.
Solving for $\psi_\beta$ in the second of $(3.19)$ and substituting
into the first,we obtain the complex Schro\"odinger type equation
$^{17}$
$$ H_{eff} \psi_\alpha = E \psi_\alpha, \eqno(3.20)$$
where
$$ H_{eff} = H_1 + H_\beta^* { 1 \over {E+H_1^*}} H_\beta
\equiv  H_1 + V_{opt}(e), \eqno(3.21)$$
and the effective Hamiltonian on the complex degrees of freedom
contains the dynamics of the quaternionic sector in the form of
an ``optical potential''.  Clearly, the quaternionic degrees of
freedom can break time reversal invariance.
\bigskip
{\bf 4. Spectral Properties and the Time Operator}
\smallskip
\par In this section, I discuss the spectral properties of
quaternion anti-self-adjoint operators, their implication
for the existence of a ``time operator'' and the possibility
of describing irreversible processes.
\par  The definition of the spectrum of ${\tilde H}$ in $(3.13)$
appears to admit of only positive spectra.  In fact, $(3.12)$
indicates that the spectrum of a quaternionic anti-self-adjoint
operator is a sphere; only the two points $\pm E$ for each $E$
will concern us now.  Multiplying $(3.13)$ by $e_2$ on the right,
one obtains
$$ {\tilde H} (\psi e_2) = \psi e_2 (-e_1 E), \eqno(4.1)$$
so that if $\psi$ is an eigenfunction with eigenvalue $E$,
$\psi e_2$ is an eigenfunction with eigenvalue $-E$.  If we
suppose an absolutely continuous spectrum for $E$ in $(0, \infty)$,
then ${\tilde H}$ has an absolutely continuous spectrum $^{19}$
in $(-\infty, \infty)$.  Discrete spectra imbedded on this
continuum (and interacting with it) may lead to resonances
with interesting properties$^{20}$, which will be discussed briefly
below.
\par   The spectal representation for an anti-self-adjoint operator
is (in the Dirac form for absolutely continuous spectrum)
$$ {\tilde H} = \int_0^{\infty} dE \vert E \rangle e_i E \langle
E \vert, \eqno(4.2) $$
so that in the sense of generalized eigenfunctions,
$$   {\tilde H} \vert E \rangle = \vert E \rangle
e_1 E. \eqno(4.3) $$
We therefore define
$$ \vert E \rangle e_2 \equiv \vert -E \rangle. \eqno(4.4)$$
Then, calling the complex part (the $\alpha$-component) of
$\langle E \vert \psi) $ by $_c \langle E \vert \psi)$, we have
$$ \eqalign{ _c\langle E \vert \psi) &= \psi_\alpha (E) \cr
_c\langle -E \vert \psi) &= \psi_\beta(E),\cr } \eqno(4.5)$$
and it follows that, with these definitions
 (in an obvious notation),
$$ {\tilde H} = \int_{-\infty}^{\infty} dE \vert E \rangle
e_1 E \ \ _c\langle E \vert, \eqno(4.6)$$
displaying explicitly the negative spectrum.  There exists an
operator $ T$ such that (in units $[ T] =
[{\tilde H}^{-1}]$)
$$      [T, {\tilde H}] = 1, \eqno(4.7)$$
where $T$ is the Hermitian operator
$$ \eqalign{ T &= -\int_0^\infty dE \vert E \rangle e_1
{ \partial \over \partial E}
\langle E \vert  \cr &= \int_{-\infty}^\infty dE
\vert E \rangle e_1 {\partial \over \partial E } \ \ {_c\langle}
E \vert . \cr} \eqno(4.8)$$
\par It is a straightforward consequence of the symmetry of
the spectrum that the quaternionic analog of the Lee-Friedrichs
model$^{21}$, based on an unperturbed Hamiltonian with a discrete
eigenvalue embedded in an absolutely continuous spectrum with
perturbation that connects the continuum only to the discrete
state , develops a complex pole below the real {\it negative}
axis as well as the usual pole below the real positive axis.  The
resulting interference term in the decay law can lead to
oscillations, and may conceivably be observable$^{20}$.
\par Misra, Prigogine and Courbage$^{21}$ have stressed the
importance of the existence of a time operator for the
description of irreversible processes.   Their demonstration
assumes that there is a Lyapunov operator (entropy) $M$ for which
$ \dot M = dM/dt \geq 0 $ and commutes with $M$.  One then easily
shows that the expectation value of the Hamiltonian in the
state $e^{iMs}\psi$, for some $\psi$ in the domain of $M,\dot M$
must be unbounded from below; it therefore
 admits a conjugate time operator.
It was partly for this reason that emphasis has been placed,
in the Brussels school, on developments of methods in the
Liouville space, where the generator of evolution (whose prototype
is the commutator with the Hamiltonian) has unbounded spectrum.
\par It has, moreover, been recently shown that the beautiful
theory of Lax and Phillips$^{22}$ for describing scattering
and resonances in hyperbolic systems is applicable in the framework
 of the quantum theory as well$^{23}$.  The semigroup property
of the evolution of an unstable system (exact exponential
decay)$^{24,25}$ can be achieved in this structure, in which
the usual Hilbert space of the quantum theory is expanded
to a direct integral of Hilbert spaces over the time axis.
The existence of a time operator is implicit in this structure,
which may be realized directly in the Liouville space$^{26,27}$,
in a relativistic quantum theory (in which the evolution operator
is also not bounded from below)$^{28}$, or in the
quaternionic Hilbert space as described above.
\bigskip
{\bf 5. Fock Space and Quantum Field Theory}
\smallskip
\par The construction of a tensor product of quaternion
modules, following the usual method for the representation
of many-body systems (and the Fock space that is the
prototype for quantum field theory) has long been an obstacle
in quaternionic theories, since there is an essential
destruction of linearity, i.e., $fq \otimes g \neq (f \otimes g)q$
, where $q \in {\bf H}$.  In his book$^{17}$, Adler has
described a new construction in the framework of path integrals
which appears to be a very powerful way of handling the problem
of the construction of a quaternionic
quantum field theory, and, in fact, of more general
 quantum field theories.  It follows, however, from the
theorem stated in $(3.9)$, that one may consider the tensor
product problem in terms of that of a real Hilbert space and
the vector space of a tensor product of quaternion algebras,
e.g., for
$$ f\otimes g = \sum_{i,j} f_i \otimes g_j \cdot e_i \otimes
e_j .  \eqno(5.1) $$
\par The first factor is relatively simple to study, since
it involves the tensor product of real Hilbert spaces.
The requirements in dealing with the second factor stem
from the fact that the tensor product space must be a
{\it quaternionic} Hilbert space, and hence the scalar
products must be quaternion valued.  Furthermore, we wish
to construct a scalar product in the direct product algebra
space which is totally symmetric, so that Bose-Einstein or
Fermi-Dirac statistics can be achieved for the physical states
by appropriate symmetrization of the $\{f,g, \dots \}$ entering
the tensor product.  This was done by Razon and me$^{29}$;
we found and studied the properties of the corresponding
annihilation-creation operators which create and annihilate
states with correct Bose-Einstein and Fermi-Dirac symmetry,
but satisfy commutation and anti-commutation realtions
which are deformed from the usual ones, i.e.,
$$ a(f) a^\dagger (g) \mp \lambda a^\dagger (g) a(f) = F(f,g),
       \eqno(5.2)$$
where $F(f,g)$ is a simple functional of $f$ and $g$, and
$\lambda$ is a real number determined by the occupation
number of the quaternionic state this relation acts on
(it may therefore be considered a function of the quaternion
number operator).  It is interesting that the scalar product
of two one-particle states  does not coincide with that of the
original Hilbert space, i.e.,
$$ \bigl(\Psi(f),\Psi(g)\bigr) = {1 \over 3} \bigl( 2(f,g)
+ (g,f) \bigr), \eqno(5.3)$$
and the annihilation of a one-particle state yields
$$ a(f) \Psi(g) = \Psi_0 {1 \over 3} \bigl( (2(f,g) + (g,f)
\bigr) .  \eqno(5.4) $$
  The reason for this is that the vacuum $\Psi_0$ carries
a non-trivial quaternionic phase, and the creatiion of a one-
particle state therefore involves the construction of a non-trivial
tensor product.  The one-particle function $\Psi(f)$ carries,
as shown in ref. 29,  a linearity property constructed for the
scalar product of N-body functionals, but not the linearity
under $f \rightarrow fq$ associated with the original space.
\bigskip
{\bf 6. Comments}
\smallskip
\par The associative Clifford algebras appear to provide
models for quantum theories generalized, in their realization,
beyond the usual complex structure.  The natural ``phase'' of
the linear spaces in these theories is non-abelian,  and they
may provide models for describing the quantum states of the
non-abelian gauge theories entering in recent attempts to
describe strong and electroweak interactions, and to account
for the observed particle spectrum.  It appears that
some fundamental structural modification should be made on the
basic form and realization of the general quantum theory if
progress is to be made in the description of these phenomena$^{17}$.
The generalization of the idea of locality, as in the theory of
strings and conformal field theories$^{30}$ is a possibility that
is being investigated widely; the modification of the algebraic
structure of the realization of quantum theory is yet another
possibility that I have discussed here.
\bigskip
{\bf References}
\smallskip
\frenchspacing
\item{1.}M. Jammer, {\it The Conceptual Development of
Quantum Mechanics,} McGraw Hill, N.Y. (1966), pp. 205, 375-377.
\item{2.}  Mr. Tait, in a letter to A. Cauchy $^1$.
\item{3.} W.R. Hamilton, Phil. Mag. {\bf 25}, 10, 241, 489 (1844).
\item{4.} H. Taber, Amer. Jour. Math. {\bf 12}, 337 (1890).
\item{5.} J.W. Gibbs, Nature {\bf 44}, 79 (1891).
\item{6.} A. Hurwitz, Nachr. Gesell. Wiss., G\"ottingen,
Math-Phys. Kl., 309 (1898).
\item{7.} Personal communication to H.H. Goldstine.  The axiomatic
foundation of the quantum theory does not restrict the
structure of the Hilbert module in which the propositional
system is embedded, provided that it is isomorphic to a
projective geometry.  See, for example, C. Piron, {\it M\'ecanique
quantique}, Presses polytechniques et universitaires romandes,
Lausanne (1990).
\item{8.} H.H. Goldstine and L.P. Horwitz, Proc. Nat. Aca. Sci.
{\bf 48}, 1134 (1962); Math. Ann. {\bf 154}, 1 (1964).
\item{9.} P. Jordan, J. von Neumann and E.P. Wigner,
Ann. Math. N.Y. {\bf 35}, 29 (1934).
\item{10.} A.A. Albert, Ann. Math. {\bf 35}, 65(1934).
\item{11.} H.H. Goldstine and L.P. Horwitz, Math. Ann.
{\bf 164}, 291 (1966).
\item{12.} M. G\"unaydin and F. G\"ursey, Phys. Rev. D{\bf 9}
, 3387 (1974).
\item{13.} L.P. Horwitz and L.C. Biedenharn, Jour. Math. Phys.
{\bf 20},  269 (1979).
\item{14.} W. Freudenthal, Math. Inst. der Rijksuniversiteit
te Utrecht (1951).
\item{15.} D. Finkelstein, J.M. Jauch, S. Schiminovitch and
D. Speiser, Jour. Math. Phys. {\bf 3}, 207 (1962);{\bf 4},
788 (1963).
\item{16.} L.P. Horwitz and L.C. Biedenharn, Ann. Phys.
{\bf 157}, 432 (1984).
\item{17.} S.L. Adler, {\it Quaternionic Quantum Mechanics
and Quantum Fields}, Oxford University Press, Oxford (1995).
\item{18.} A. Razon,L.P. Horwitz and L.C. Biedenharn, Jour.
Math. Phys. {\bf 30}, 59 (1989).
\item{19.} L.P. Horwitz, Jour. Math. Phys. {\bf 34}, 3405 (1993).
\item{20.} L.P. Horwitz, Jour. Math. Phys. {\bf 35}, 2743,2760
 (1994).
\item{21.} T.D. Lee, Phys. Rev. {\bf 95}, 1329 (1954); K.O.
Friedrichs, Comm. Pure and Appl. Math. {\bf 1}, 361 (1950).
\item{22.} B. Misra, I. Prigogine and M. Courbage, Proc.
Nat. Aca. {\bf 76}, 4768 (1979).
\item{24.} C. Flesia and C. Piron, Helv. Phys. Acta {\bf 57}, 697
(1984).
\item{25.} L.P. Horwitz and C. Piron, Helv. Phys. Acta {\bf 66},
693 (1993).
\item{26.} E. Eisenberg and L.P. Horwitz, Adv. Chemical Phys.,
to be published.
\item{27.} E. Eisenberg and L.P. Horwitz, Phys. Rev. A {\bf 52},
70 (1995).
\item{28.} Y. Strauss and L.P. Horwitz, in preparation.
\item{29.} A. Razon and L.P. Horwitz, Acta. Appl. Math. {\bf 24},
141, 179 (1991).
\item{30.} For example, M.B. Green, J.H. Schwarz and E. Witten,
{\it Superstring Theory} I and II, Cambridge Univ. Press,
Cambridge (1987).

\vfill
\eject
\end
\bye